# Next-Generation 6G Networks: Deploying Cybertwin Technology for Enhanced Healthcare Solutions

Alinafe Kaliwo
*Department of Computer Science*
*University of the Western Cape*
Cape Town, South Africa
akaliwo@mubas.ac.mw

Clement Nyirenda
*Department of Computer Science*
*University of the Western Cape*
Cape Town, South Africa
cnyirenda@uwc.ac.za

*Abstract*—This paper explores the integration of Cybertwin technology within 6G networks to revolutionize healthcare delivery. It aims to enhance real-time monitoring, decision-making, and resource management through the Service-based Hierarchical Framework for Cybertwins in sixth-generation networks. The paper addresses the deployment challenges and proposes system theory as a comprehensive framework for designing complex interactions among healthcare-assigned Cybertwins. The article highlights the role of Cybertwin technology in advancing healthcare solutions, promising improved patient care and operational efficiency. It examines the current network setups and the potential of 6G infrastructure, discussing network topology optimization, theoretical modelling, and future directions. The paper underlines the transformative impact of combining 6G and Cybertwin technologies on healthcare, from high-definition telemedicine to large-scale patient monitoring. The paper further acknowledges the implementation challenges, such as technical complexity, security, and interoperability.

## I. INTRODUCTION

The advent of sixth-generation (6G) technology is expected to bring about a revolutionary transformation in healthcare by providing better connectivity, faster speed, and increased reliability. This next-generation network technology is designed to enable further-enhanced mobile broadband (FeMBB), extremely reliable and low latency communications (ERLLC), long-distance and high mobility communications (LDHMC), ultra-massive machine-type communications (umMTC), and extremely low-power communications (ELPC) [1], leading to advancements such as real-time remote patient monitoring, telehealth services, and AI-powered diagnostics [2]These capabilities promise to make healthcare more accessible, efficient, and personalized, thus significantly impacting patient outcomes and healthcare delivery models.

Moreover, 6G technology's potential to support edge computing will allow for faster health data processing and analysis at or near the point of care. This will considerably reduce the dependence on central data centres and minimize delays, which is crucial for critical healthcare applications such as telesurgery, holographic communications, augmented and virtual reality, five-sense communications, and tactile internet, where every millisecond counts [1]. By providing the backbone for innovative healthcare solutions, 6G technology will play a pivotal role in shaping the future of medical care and public health [3].

In the context of 6G networks, Cybertwin technology represents a transformative approach to healthcare by offering a more advanced combination of digital modelling and simulation technologies [4] for real-time monitoring, decision-making, and resource management. These digital-physical twins, precise virtual models of physical entities, can simulate the conditions of patients, medical devices, or entire healthcare systems in real-time. This capability enables healthcare providers to anticipate patient needs, optimize treatments, and manage resources more efficiently, enhancing patient care and operational effectiveness.

By leveraging the extremely reliable and low-latency communication features of 6G, Cybertwin technology ensures instantaneous data synchronization between the physical and virtual worlds, facilitating immediate responses to medical emergencies and enabling predictive healthcare analytics. The integration of Cybertwins with 6G networks promises to unlock new levels of precision in patient monitoring, personalized care, and system resilience, paving the way for a reactive healthcare ecosystem that is predictive and adaptive to changing patient needs [5]. The potential of 6G technology and Cybertwin technology to enhance healthcare is immense, and it is expected to lead to significant improvements in patient outcomes and healthcare delivery models.

This paper explores various crucial themes to address the complexities and emerging potential of integrating Cybertwin technology within 6G networks for healthcare. We delve into the substantial body of related work on Cybertwins and 6G, detailing their roles in network management, optimization, and security, alongside notable contributions to the field. The foundational aspects of Cybertwin technology, from its aerospace origins to its evolution into a tool for personalization and interaction through AI and machine learning, are examined. Further, we discuss the integration of Cybertwins in 6G networks, highlighting their functionality in real-time optimization and the architectural nuances enabling these capabilities. Our exploration extends to design considerations critical for deploying Cybertwins in 6G healthcare environments, including reliability, security, and AI-driven analytics. We propose a comprehensive framework—the *Service-based Hierarchical Framework for Cybertwins in 6G (SHFC6G)*—to underpin our architectural and theoretical discussions. Implementation challenges, spanning technical, regulatory, and user acceptance aspects, are analyzed to offer a holistic view of the deployment intricacies. Lastly, we project future directions, emphasizing AI, machine learning advancements, and the need for cross-disciplinary collaboration to navigate regulatory and economic models. Through this exploration, we aim to provide a well-rounded understanding of deploying Cybertwins in 6G-enabled healthcare systems, their potential impacts, and the hurdles to their realization.

## II. RELATED WORK

The usage and integration of Cybertwins in communication networks have been extensively researched over the past few decades. This section provides a brief overview of the most significant works and recent developments in this field, particularly in the context of emerging 6G and Cybertwin technologies.

### A. Integration of Cybertwins in Communication Networks

In communication networks, Cybertwins serve as virtual counterparts that enhance network management and optimization. These digital replicas facilitate critical tasks, including modelling network behaviour, simulating scenarios, and testing "what-if" situations. By leveraging Cybertwins, network operators can fine-tune configurations, optimize performance, and ensure efficient operation of networks [6]. Additionally, Cybertwins play a pivotal role in product security, aiding experts in identifying vulnerabilities, analyzing zero-day threats, ensuring compliance, and strengthening supply-chain security. In summary, Cybertwins empower network evolution, security, and automation, shaping the future of interconnected systems.

The integration of Cybertwins into communication networks has been a captivating novel area of research. In one of the notable research papers [7], authors proposed a Cybertwin-assisted joint mode selection and dynamic pricing (JMSDP) scheme for effective network management in ultra-dense low earth orbit (LEO) integrated satellite-terrestrial networks (ULISTN). In their setup, the Cybertwin is an intelligent agent, assisting in mode selection decisions. The paper discusses optimal access prices for terrestrial small base stations (TSBSs) and terrestrial-satellite terminals (TSTs), enabling IoT users to select their access mode based on these prices. The proposed JMSDP improves average throughput and reduces delay compared to random access (RA) and maximum rate access. Future research opportunities include designing reference data management architectures, adopting industry standards and ontologies, ensuring interoperability among distinct Cybertwins, and implementing robust data provenance mechanisms.

In another paper, the authors propose a groundbreaking network architecture called Cybertwin-Based Cloud Native Network (CCNN). To address the ensuing challenges in the future Internet of Everything (IoE), including network heterogeneity, application cloudification, and personalized user services, the CCNN merges three critical components: the radio access network (RAN), the IP bearer network, and the data centre network. Key features of CCNN include unified virtualization using Kubernetes, a decoupled RAN architecture for efficient resource utilization, and an immunology-inspired security framework combining zero-trust security principles with an adaptive defence system [8].

In another paper titled [9], authors propose a novel mechanism that dynamically selects and adjusts transmission control methods in software-defined vehicular networks. Leveraging deep reinforcement learning (DRL) and Cybertwin-driven approaches, their solution, called TcpCDRL, outperforms traditional methods regarding network throughput and round-trip time (RTT). This research contributes valuable insights to the field of vehicular networks.

### B. Cybertwins in 5G Networks and Beyond

The integration of Cybertwins into fifth-generation and beyond (B5G) networks is crucial in improving network management, security, and optimization. Cybertwins have the potential to be used in various applications, such as intelligent traffic prediction, data management, and predictive maintenance, which can revolutionize the way interconnected systems are managed and operated.

The authors of [10] propose a framework that integrates Cybertwins with 6G-enabled edge networks. This framework focuses on intelligent service provisioning for various Internet of Everything (IoE) applications. The authors achieve significant improvements in energy efficiency and prediction accuracy by using deep reinforcement learning to distribute tasks dynamically and utilizing an artificial intelligence-driven technique based on the support vector machine (SVM) classifier model. This work provides valuable insights into the role of Cybertwins in shaping resource allocation within 6G edge networks.

In another paper [11], authors propose a novel network virtualization (NV)-based architecture for Cybertwin-enabled 6G core networks. The central challenge lies in optimizing the virtual network (VN) topology—composed of virtual nodes and intermediate virtual links—and determining the resultant VN embedding within a Cybertwin-enabled substrate network. The authors formulate an optimization problem to minimize the embedding cost while meeting end-to-end (E2E) packet delay requirements. They leverage queueing network theory to evaluate each service's E2E packet delay, which depends on the resources allocated to the virtual nodes and links. The problem is formally a mixed-integer nonlinear program (MINLP), and the authors propose an improved brute-force search algorithm to find optimal solutions. An adaptively weighted heuristic algorithm is also introduced for large-scale networks, yielding near-optimal solutions. Simulations demonstrate the effectiveness of these algorithms in enhancing network performance compared to benchmark approaches. The study contributes valuable insights into 6G, Cybertwin, and virtual network embedding (VNE).

In yet another research [12], researchers explore the concept of a muscular human Cybertwin—a novel extension beyond traditional artificial neural networks (ANNs) and musculoskeletal models (MSMs). The Cybertwin-driven 6G technology, capable of obtaining static and dynamic data streams from users, opens up exciting possibilities. The proposed conceptual design combines ANN and MSM, leveraging learning-based approaches and analytical methods. The resulting baseline model demonstrates improved generalization ability over ANN and easier adaptation to new data distributions than MSMs. The study evaluates joint moment prediction accuracy, data efficiency, generalization, and personalization time efficiency. Remarkably, the method achieves accuracy similar to ANN while outperforming MSM by over 30% with sufficient training data. Furthermore, it exhibits over 70% accuracy improvements in unseen walking conditions and new subjects. The time efficiency during model fine-tuning is also significantly better than calibrated MSM.

## III. Technology Overview

### A. Foundations of Cybertwin Technology

The concept of a "Cybertwin" has its roots from the digital twin technology, originating from the fields of aerospace by the National Aeronautics and Space Administration (NASA) in the late 1960s in the Appollo 13 mission [13], [14] and manufacturing in the early 2000s. Digital twin technology was initially used to create detailed digital replicas of physical objects or systems for simulation, analysis, and optimization purposes. The idea was to have a virtual counterpart that could mirror the real-world entity in every aspect, allowing for better understanding, predictive maintenance, and innovation without the risk of damaging the actual object [15]. Over time, this technology evolved beyond industrial applications to encompass more personal and interactive uses, leading to the development of Cybertwins that mimic human behavior, preferences, and interactions. This evolution was significantly propelled by advancements in artificial intelligence, machine learning, and data analytics, enabling more sophisticated and personalized digital entities that can interact with users in various contexts, from customer service bots to personalized avatars and virtual assistants [16].

The potential of Cybertwins goes beyond mere personalization and interaction. With the ability to learn from user behavior and preferences, Cybertwins can offer tailored recommendations and solutions that meet the specific needs of individuals. They can also be used to create virtual simulations of complex scenarios, such as disaster response or emergencies, to help improve preparedness and response. Moreover, using Cybertwins in healthcare is promising, as they can learn from patients' medical history, symptoms, and treatment responses to offer personalized healthcare solutions.

As the technology behind Cybertwins advances, we can expect to see more innovative uses and applications in various industries. The possibilities are endless, and the benefits of using Cybertwins are undeniable. They offer a safe, cost-effective, and efficient way to simulate, analyze, and optimize real-world scenarios without the risk of damaging the actual object.

### B. Integration of Cybertwins in 6G Networks

Within the adaptive and advanced architecture of 6G, as illustrated in Fig. 1, Cybertwins promise to serve as virtual replicas of network components, user devices, and services, leveraging the network's artificial intelligence (AI) and machine learning capabilities for real-time optimization, predictive maintenance, and personalized user experiences [17].

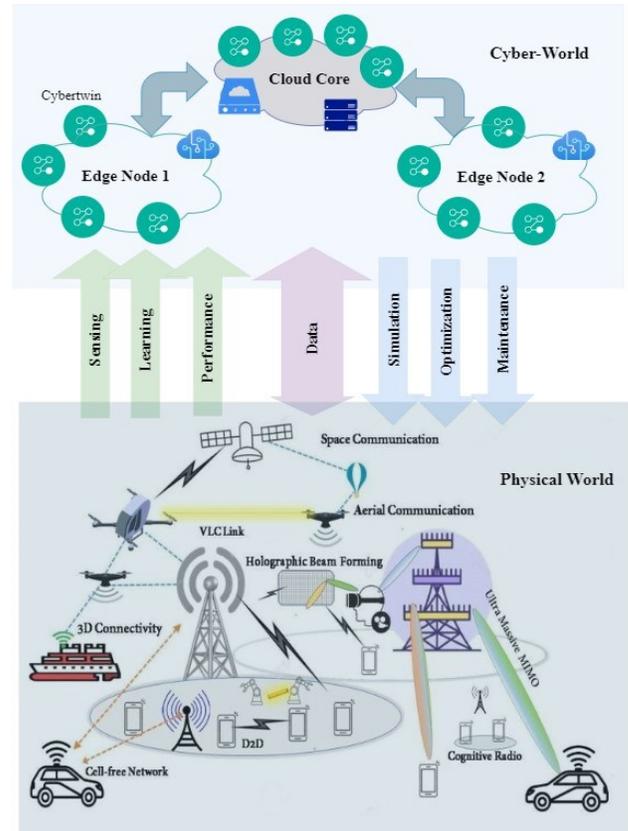

Fig. 1. Interlinkages between the 6G physical- and cyber- worlds using Cybertwins (adapted from [18], [19]).

By sensing and simulating the behavior and performance of physical entities in the network, Cybertwins will enable proactive adjustments to network configurations, enhancing efficiency, reducing latency, and ensuring reliability for a wide array of applications, from immersive augmented reality (AR) experiences to critical Internet of Things (IoT) deployments [18]. The use of high-frequency bands and edge computing in 6G architecture further supports the seamless operation of Cybertwins by providing the necessary speed and data processing capabilities near the user, thus minimizing response times and enabling more sophisticated interactions.

## IV. Design Considerations in 6G Healthcare Networks

### A. Based on the Capablities of 6G Networks

The deployment of Cybertwins in 6G healthcare systems necessitates careful consideration of several key design principles. Firstly, the system must ensure robust and reliable connectivity, given that 6G networks promise extremely reliable and low latency communications (ERLLC). This is crucial for real-time health monitoring and emergency medical response.

Secondly, the system should prioritize data privacy and security. As Cybertwins handle sensitive health data, robust encryption and secure data transmission protocols are essential to protect patient information.

Furthermore, the system should be designed for seamless interoperability with various healthcare devices and systems. This includes compatibility with wearable devices, medical imaging systems, and electronic health record systems. Lastly, the system should incorporate advanced AI algorithms for predictive analytics, enabling proactive health management. The Cybertwins, powered by AI, can analyze health data in real time, predict potential health risks, and provide personalized health recommendations, thereby transforming the healthcare experience in the 6G era.

### B. Based on Key Standards, Requirements and Specifications

Designing Cybertwins for 6G networks in healthcare involves several key considerations and standards. The upcoming 6G communication systems are expected to provide Intelligent Internet of Healthcare Things (IIoHT) services everywhere to improve the quality of life. This includes massive and smart connectivity, huge bandwidth, lower latency with ultra-high data rate, and better quality of healthcare experience. The 6G cellular network should integrate with various communication techniques such as optical wireless communication network, cell-free communication system, backhaul network, and quantum communication. A distributed security model is essential in the context of IIoHT to ensure the safety and privacy of health data. Furthermore, AI technologies are crucial for creating a robust infrastructure with low latency and ultra-high-speed access networks.

The future healthcare system demands high data rate ($\geqslant 1$ Tbps), high operating frequency ($\geqslant 1$ THz), low end-to-end delay ($\leqslant 1$ ms), high reliability ($10^{-9}$), high mobility ($\geqslant 1000$ km/h), and wavelength of $\leqslant 300$ μm. Intelligent Wearable Devices (IWD) and Intelligent Internet of Medical Things (IIoMT) will play a significant role in the healthcare system of the 6G era. 6G will enable robust Hospital-to-Home (H2H) services, enhancing the quality of life and making healthcare more accessible. Lastly, 6G communication technology will enable real-time remote surgeries.

## V. Proposed Framework

### A. Service-based Hierarchical Framework of Implementing Cybertwins in 6G Networks (SHFC6G)

We propose a slice-based hierarchical system for healthcare applications that leverages the advanced capabilities of 6G technologies to create a multi-layered, interconnected network of Cybertwins called the SHFC6G. At the core of the SHFC6G system are individual slice or service-based Cybertwins, digital replicas of physical entities representing a group of patients, medical devices, services or even entire healthcare facilities. These Cybertwins can continuously collect and analyze data in real time, providing valuable insights and enabling proactive service-based healthcare interventions. The high-speed and low-latency characteristics of 6G networks ensure that these Cybertwins can communicate and synchronize with their physical counterparts almost instantaneously as they share data, allowing for real-time monitoring and decision-making.

The hierarchical structure of the system allows for efficient data management and scalability. At the lower levels of the hierarchy, Cybertwins handle more granular tasks, such as monitoring specific patient's vital signs, connected service or managing a single medical device. As we move up the hierarchy, the Cybertwins take on more complex tasks that require a broader view, such as coordinating the operations of an entire hospital ward or managing a city's healthcare resources. This structure allows for efficient data flow and decision-making, with each level of the hierarchy focusing on its specific tasks while also contributing to the overall goals of the system. The massive data handling capabilities of 6G networks make it possible to manage and analyze the vast amounts of data generated by the Cybertwins, leading to more informed and effective healthcare decisions.

### B. Architectural Representation of SHFC6G for Healthcare Solutions

In healthcare services and solutions, the SHFC6G assigns specific healthcare services to Cybertwins at each level of the hierarchy. These services can then be grouped in slices, which can then be monitored and enhanced by slice-based Cybertwins, as below.

*1) Further-Enhanced Mobile Broadband Cybertwin (FeMBB$_C$):* For high-definition telemedicine consultations, where doctors and patients interact in real-time with high-quality video.

*2) Extremely Reliable and Low Latency Communications Cybertwin (ERLLC$_C$):* For remote surgeries or teleoperations where a surgeon controls robotic instruments from a distance.

*3) Long-Distance and High Mobility Communications (LDHMC$_C$) Cybertwin:* For ambulance services, where patient data can be transmitted to the hospital while the ambulance is en route.

*4) Ultra-Massive Machine-Type Communications Cybertwin (umMTC$_C$):* For large-scale remote monitoring of patients, where thousands of wearable devices transmit health data to healthcare providers.

*5) Extremely Low-Power Communications Cybertwin (ELPC$_C$):* For long-term remote patient monitoring using wearable or implantable devices.

In Fig. 2, the hierarchical implementation of Cybertwins within a 6G network for healthcare is depicted. The subscript 'C' identifies the components as Cybertwins, whereas '*i*' and '*g*' distinguish between individual and global Cybertwins, depending on their level within the hierarchy.

At the base, individual Cybertwins ('i') at the edge nodes are focused on specific tasks, like monitoring a single patient's health metrics or managing a particular medical device. These are the frontline units of digital representation that interact directly with the physical elements they represent.

As we move up the hierarchy, global Cybertwins ('g') at the edge node level are responsible for synthesizing information from the individual Cybertwins. They might oversee the functions of various medical devices and patient data within one healthcare facility, ensuring the local network's operational coherence and immediate responsiveness.

At the highest tier in the cloud native core, the global Cybertwins ('g') handle comprehensive tasks. These Cybertwins aggregate data and functionalities from across all edge nodes, allowing for integrated and strategic management of an entire healthcare system or even city-wide healthcare resources.

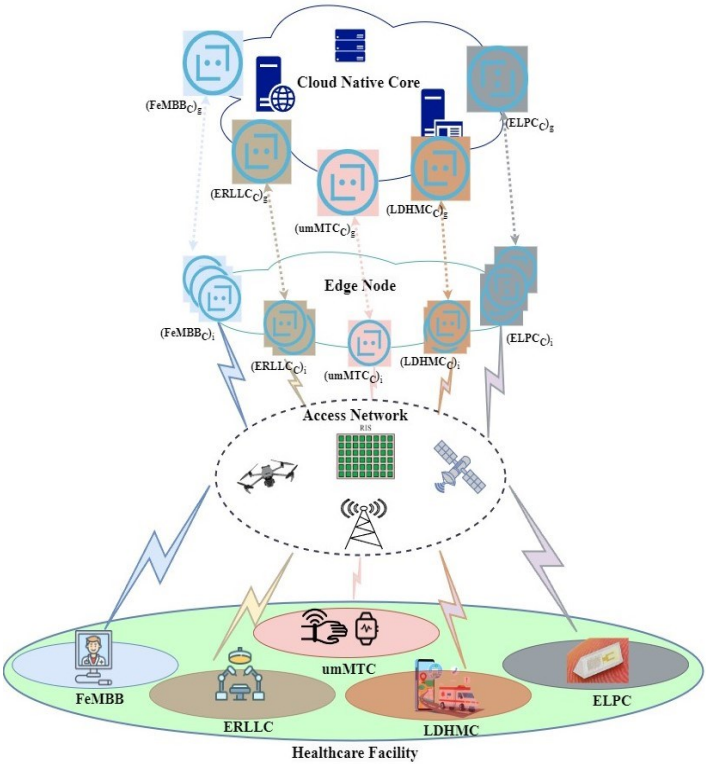

Fig. 2. Hierarchical architecture of Cybertwins in a 6G-enabled healthcare network.

This system utilizes the advanced capabilities of 6G networks to facilitate real-time communication and data processing across various levels of Cybertwins, optimizing healthcare delivery and resource allocation at scale.

### C. Proposed Protocol Stack for the Integration of Cybertwins in 6G Networks for Optimal Service Delivery

In the proposed Cybertwin framework for 6G-enabled healthcare networks, a meticulously designed protocol stack ensures seamless, secure, and efficient communication between node components. At the application layer, a specialized Application Layer Protocol (ALP) facilitates Cybertwin interactions, supporting healthcare applications directly within edge and core network nodes. The Message Queuing Telemetry Transport (MQTT) protocol, positioned at the session layer, offers a lightweight messaging system for IoMT devices, bridging the application layer to the transport protocols. For securing data transmission, the Transport Layer Security (TLS) protocol is employed at the transport layer, ensuring end-to-end encryption of sensitive healthcare data. The stack also incorporates the User Datagram Protocol (UDP) and Quick UDP Internet Connections (QUIC) for low-latency communications critical for real-time applications, while IPv6 at the network layer addresses the expansive needs of a 6G network's addressing and routing. Finally, at the foundation, Physical and Data Link Protocols, such as 5G NR and emerging 6G standards, are responsible for the physical transmission of data, ensuring that the infrastructure supports the high-bandwidth, low-latency requirements of advanced healthcare services. This holistic approach to the protocol stack across different layers guarantees that the Cybertwin framework leverages 6G capabilities to deliver transformative healthcare solutions efficiently and securely.

### D. Proposed Theory of Implementation of Cybertwins in the SHFC6G

**System Theory** is proposed to provide a general framework that could be instrumental in understanding and designing the complex interactions between the various Cybertwins assigned to healthcare services. By embracing a holistic viewpoint, system theory stresses the importance of viewing the network as more than just the sum of its parts, focusing on how individual Cybertwins at various levels—ranging from individual services like high-definition telemedicine (using $FeMBB_C$) to the global management of city-wide healthcare resources.

This approach includes ensuring seamless interconnectivity for efficient data flow, integrating feedback mechanisms for adaptability, and building redundancy for resilience, ensuring that critical healthcare services are maintained without interruption. Moreover, system theory examines the optimization of the entire network, considering aspects such as resource allocation, energy consumption, and service quality to enhance overall performance and efficiency. We are still working on relevant mathematical models to support the system theory principles, and one of the concepts was presented in a paper that we published [5].

## VI. IMPLEMENTATION CHALLENGES

Implementing the SHFC6G framework with its hierarchy of Cybertwins introduces significant technical challenges, primarily in designing a network capable of fulfilling diverse healthcare service requirements such as high-bandwidth telemedicine and efficient, low-power patient monitoring. The network demands a seamless interplay of Cybertwins, ensuring secure data management, interoperability between disparate devices, and real-time response capabilities. Additionally, it must be resilient and reliable, capable of recovering swiftly from various disruptions, including hardware, software, or security breaches, which is paramount given the critical nature of healthcare services.

Beyond technical hurdles, the network's deployment confronts practical challenges like adhering to stringent healthcare regulations, achieving economic viability to justify the investments, and ensuring scalability to accommodate an expanding ecosystem of devices and data. Equally important is gaining user acceptance; healthcare providers and patients must trust and embrace this technology, necessitating comprehensive training and a user-friendly interface. These multifaceted challenges demand a collaborative approach that merges technological advancement with strategic policy development, compliance with healthcare laws, and a commitment to user-centric design.

## VII. Future Directions and Conclusion

As the SHFC6G network continues to evolve, future directions will likely focus on advancing the integration of AI and machine learning to enhance the autonomous capabilities of Cybertwins, improving predictive analytics for healthcare outcomes, and personalizing patient care. Research and development efforts may also concentrate on refining 6G technologies to increase network capacities, reduce latency even further, and introduce new services that could revolutionize remote healthcare. Innovations in battery technology and energy harvesting could extend the life of wearables and implanted devices, which is critical for Extremely Low-Power Communications Cybertwins. Additionally, cross-disciplinary collaborations will be essential to navigate the complexities of regulatory landscapes, develop new economic models for healthcare technology, and foster user-centric design that prioritizes ease of use and public trust.

In conclusion, implementing a hierarchical network of Cybertwins in a 6G healthcare framework holds immense potential to transform the healthcare industry. It promises to enhance the quality of care, improve the efficiency of healthcare delivery, and enable new forms of remote health monitoring and intervention. While there are significant challenges to overcome, particularly in terms of system complexity, security, and user acceptance, the continued refinement and adaptation of the SHFC6G network position it as a pivotal component in the future of healthcare technology. The successful realization of this vision will depend on a concerted effort from technology developers, healthcare professionals, policymakers, and the community at large.

## Acknowledgement

The authors would like to express their gratitude to everyone who contributes to this work, particularly the researchers from the Computer Science Department of the Faculty of Natural Science at the University of the Western Cape. This research was supported by the South Africa National Research Foundation (NRF) Grant No. 980000001090, and Telkom South Africa, through the Telkom Centre of Excellence (CoE) at the University of the Western Cape. The authors would also like to acknowledge the support from the Malawi University of Business and Applied Sciences. The authors take full responsibility for the statements made and views expressed in this work.